\documentstyle[12pt]{article}
\begin{document}
\pagestyle{plain}
\title{\bf The propagation of a pulse in the real strings and rods}
\author{ \bf Miroslav Pardy \\
Department of Physical electronics and  \\
the laboratory of plasma physics  \\
Masaryk University \\
Kotl\'{a}\v{r}sk\'{a} 2, 611 37 Brno, Czech Republic\\
email:pamir@physics.muni.cz}
\date{\today}
\maketitle

\begin{abstract}
We consider the elastic rod of a large  mass $M$, the
left end of which is fixed  to a body of mass $m << M$
and the second body of mass $m$ is fixed to the right end of the rod.
The force of the delta-function form is applied to the left side of the rod.
We find the propagation of the pulse in the system.
Our problem represents the missing problem in the
Newton  ``Principia mathematica'' and in any textbook on mechanics.
 The relation of our theory to the quark-string model of mesons is evident.
\end{abstract}

\vspace{10mm}
\section{Introduction}
\hspace{3ex}

It is interesting to consider the elastic rod of a large  mass $M$, the
left end of which is joined with mass $m << M$
and body of mass $m$ is fixed to the right end of the rod. Then, 
it is interesting to study the consequences of the application of the 
the force of the delta-function form to the left side of the rod.
 The delta-function is chosen for simplicity. This
function can be replaced by the different functions. We show that the internal
motion of the elastic rod medium is controlled by the wave equation. We derive 
the mathematical form of the mechanical motion of the considered string or rod.
Our problem represents
the missing problem in the Newton  ``Principia mathematica'' [1]
and in any textbook on mechanics. The relation of our theory to the quark-string model of mesons is evident.

\section{Classical theory of interaction of particle with an  impulsive force}

We will first show that use of the impulsive force of the delta-function form
is physically meaningful in a classical mechanics of a point particle.
We idealize the impulsive force by the Dirac $\delta$-function.

Newton's second law in the one-dimensional form for the interaction
of a massive particle with mass $m$ with force $F$

$$ma = F\eqno(1)$$
with $F$ being an impulsive force $P\delta(t)$ is as follows:

$$m\frac {d^{2}x}{dt^{2}} = P\delta(\alpha t),\eqno(2)$$
where $P$ and $\alpha$ are  some constants, with MKSA dimensionality [$P$]
= ${\rm kg.m.s}^{-2}$, [$\alpha$] = ${\rm s}^{-1}$. We put $|\alpha| = 1$.

Using the Laplace transform [2] in the last equation, with

$$\int_{0}^{\infty}e^{-st}x(t)dt \stackrel{d}{=} X(s), \eqno(3)$$

$$\int_{0}^{\infty}e^{-st}{\ddot x}(t)dt = s^{2}X(s) - sx(0) - {\dot x}(0),
\eqno(4)$$

$$\int_{0}^{\infty}e^{-st}\delta(\alpha t)dt = \frac{1}{\alpha}, \eqno(5)$$
we obtain:

$$ms^{2}X(s) - msx(0) - m\dot x(0) = P/\alpha.\eqno(6)$$

For a particle starting from the rest with $\dot x(0) = 0, x(0) = 0$, we get

$$X(s) = \frac{P}{ms^{2}\alpha}.\eqno(7)$$
Using the inverse Laplace transform, we obtain

$$x(t) = \frac{P}{m\alpha}t\eqno(8)$$
and

$$\dot x(t) = \frac{P}{m\alpha}.\eqno(9)$$

In case of the harmonic oscillator with the damping force and under
influence of the general force $F(t)$, the Newton law is as follows:

$$m\frac {d^{2}x(t)}{dt^{2}}  + b{\dot x}(t)
+ kx(t) = F(t).\eqno(10)$$

After application of the Laplace transform (3) and with regard to the same
initial conditions as in the preceding situation,
$\dot x(0) = 0, x(0) = 0$, we get the following algebraic equation:

$$ms^{2}X(s) +bsX(s) + kX(s) = F(s),\eqno(11)$$
or,

$$X(s) = \frac{F(s)}{m\omega_{1}}
\frac {\omega_{1}}{(s + b/2m)^{2} + \omega_{1}^{2}} \eqno(12)$$
with $\omega_{1}^{2} = k/m - b^{2}/4m^{2}$.

Using inverse Laplace transform denoted by symbol ${\cal L}^{-1}$ applied
to multiplication of functions $f_{1}(s)f_{2}(s)$,

$${\cal L}^{-1}(f_{1}(s)f_{2}(s)) = \int_{0}^{t} d\tau
F_{1}(t-\tau)F_{2}(\tau),
\eqno(13)$$
we obtain with $f_{1}(s) =  F(s)/m\omega_{1}, \quad f_{2}(s) = \omega_{1}/
((s + b/2m)^{2} + \omega_{1}^{2}), \quad F_{1}(t) = F(t)/m\omega_{1},\\
F_{2}(t) = \exp{(-bt/2m)}\sin\omega_{1}t$.

$$x(t) = \frac {1}{m\omega_{1}}\int_{0}^{t}F(t-\tau)e^{-\frac {b}{2m}\tau}
\sin(\omega_{1}\tau)d\tau.\eqno(14)$$

For impulsive force $F(t) = P\delta(\alpha t)$, we have from the last formula

$$x(t) = \frac {(P/\alpha)}{m\omega_{1}}e^{-\frac {b}{2m}t}\sin\omega_{1}t.
\eqno(15)$$

\section{The pulse propagating in a rod}
\hspace{3ex}

In this section we will solve the motion of a string or rod with the massive
ends (the body with mass $m$ is fixed to the every end of the string)
on the assumption that the tension in the string is linear and the
applied force is of the Dirac delta-function.
First, we will derive the Euler wave equation from the Hook law of tension
and then we will give the
rigorous mathematical formulation of the problem.  Linearity of the
wave equation enables to solve this problem  by the Laplace transform method.
We follow [3] and the author preprint [4] where this method was used
to solve the Gassendi model of gravity. Although Gassendi [5] is
known in physics as the founder of the modern atomic theory of matter,
his string model of gravity was not accepted. The Newton reaction to this
model was empirical. He said: ``Hypotheses non fingo''. It seems that
Gassendi ideas was applied later by Faraday in his theory of
electromagnetism. We know also that Gassendi was independent thinker
and he was persecuted. Every
independent thinker is persecuted in any society.

The present problem can be also defined as a central collision of two
bodies (balls). While in the basic mechanics the central collision is
considered as a contact collision of the two balls, here,
the collision is mediated by the string, or, rod.

To our knowledge,  the present problem is not involved
in the textbooks of mathematical physics or in the mathematical journals.
This problem was not possible to define and solve in the Newton
period, because the method of solution is based on  the
Euler partial wave equation,
the Laplace transform, The Riemann-Mellin transform, the Bromwich
integral and Bromwich contour and other ingredients of the operator calculus
which was elaborated after the Newton period. So, this is why the
problem is not involved in the Newton ``Principia Mathematica'' [1].

Now, let us consider the rod (or string) of the length $L$,
the left end of which is joined with mass $m$ and
the right end is joined with mass $m$. The force
of the delta-function form is applied to the left end
and  the initial state of the rod is the sate of equilibrium.
The deflection of the rod  element $dx$ at point $x$ and time $t$ let
be $u(x,t)$ where $x\in(0,L)$.

The differential equation of motion of string elements can be derived
by the following way [3]. We suppose  that the force acting on
the element $dx$ of the string is given by the law:

$$T(x,t) = ES\left(\frac {\partial \*u}{\partial x}\right), \eqno(16)$$
where $E$ is the modulus of elasticity, $S$ is the cross section of
the string. We easily derive that

$$T(x+dx)-T(x) = ESu_{xx}dx. \eqno(17)$$

The mass $dm$  of the element $dx$ is $\varrho ESdx$,
where $\varrho = const$ is the mass
density of the string matter and the dynamical equilibrium gives

$$\varrho\*Sdx u_{tt} = ESu_{xx}dx. \eqno(18)$$
So, we  get

$$\frac {1}{c^2}u_{tt} - u_{xx} = 0; \quad
c  = \left(\frac {E}{\varrho}\right)^{1/2}. \eqno(19)$$

Now, we get the problem of the mathematical physics in the form:

$$u_{tt} = c^{2}u_{xx}\eqno(20)$$
with the initial conditions

$$u(x,0) = 0; \quad u_{t}(x,0) = 0\eqno(21)$$
and with the boundary conditions

$$mu_{tt}(0,t) =  au_{x}(0,t) + P\delta(\alpha t);\quad
mu_{tt}(L,t) =  au_{x}(L,t),\eqno(22)$$
where we have put

$$a = - ES;\quad P = {\rm some\; constant}.\eqno(23)$$

The delta-function can be approximatively realized by the strike of
the hammer to the left end of the rod.

The equation (20) with the initial and boundary conditions (21) and (22)
represents one of the standard problems of the mathematical physics and can
be easily solved using the Laplace transform [2]:

$$\hat L u(x,t) \stackrel{d}{=}\int_{0}^{\infty}e^{-pt}
u(x,t)dt \stackrel{d}{=} \varphi(x,p). \eqno(24)$$

Using (24) and (20) we get:

$$\hat Lu _{tt}(x,t) = p^{2}\varphi(x,p) - pu(x,0) - u_{t}(x,0) =
p^{2}\varphi(x,p),\eqno(25)$$

$$\hat L u_{xx}(x,t) =  \varphi_{xx}(x,p);\quad
\hat L \delta(\alpha, t) = 1/\alpha .
\eqno(26)$$

After elementary mathematical operations we get the differential
equation for $\varphi$ in the form

$$\varphi_{xx}(x,p) - k^{2}\varphi(x,p) = 0; \quad
k = p/c. \eqno(27)$$
with the boundary condition in eq. (22).

We are looking for the the solution of eq. (27) in the form

$$\varphi(x,p) =c_{1}\cosh k x + c_{2}\sinh k x .\eqno(28)$$

We get from the boundary conditions in eq. (22)

$$c_{1} = \frac {1}{p}\;\frac {ac(P/\alpha) \cosh (pL/c) -
(P/\alpha)mpc^2\sinh (pL/c)}
{\sinh(pL/c)(a^2 - m^2 p^2 c^2)},\eqno(29)$$

$$c_{2} = -\frac {(P/\alpha)c}{ap} + \frac {(P/\alpha)mac^2\cosh(pL/c) -
(P/\alpha)pm^2c^3\sinh (pL/c)}{a\sinh (pL/c)
(a^2 - m^2 c^2 p^2)}.\eqno(30)$$

The corresponding $\varphi(x,p)$ is of the form:

$$\varphi(x,p)= \frac {1}{p}\;\frac {ac(P/\alpha)\cosh (pL/c) -
(P/\alpha)mpc^2\sinh(pL/c)}
{\sinh (pL/c)(a^2 - m^2 p^2 c^2)}\cosh (px/c)\quad + $$

$$\left[-\frac {(P/\alpha)c}{ap} +
\frac {a(P/\alpha)mc^2\cosh(pL/c) -
bpm^2c^3\sinh (pL/c)}{a\sinh (pL/c)(a^2 - m^2 c^2 p^2)}\right]\sinh(px/c).
\eqno(31)$$

The corresponding function $u(x,t)$  follows from the theory of the Laplace
transform as the mathematical formula (res is residuum)[2]:

$$u(x,t) = \frac {1}{2\pi i}\oint e^{pt}\varphi(x,p) dp =
\sum_{p=p_{n}}{\rm res}\;e^{pt}\varphi(x,p) = $$

$$\sum _{p=p_{n}}{\rm res}\;e^{pt}
\frac {1}{p}\;\frac {ac(P/\alpha) \cosh (pL/c)}
{\sinh (pL/c)(a^2 - m^2 p^2 c^2)}\cosh (px/c) \quad -$$

$$\sum _{p=p_{n}}{\rm res}\;e^{pt}
\frac {(P/\alpha)mc^2}
{(a^2 - m^2 p^2 c^2)}\cosh (px/c)\quad - $$

$$\sum _{p=p_{n}}{\rm res}\;e^{pt}
\left[\frac {(P/\alpha)c}{ap}\right]\sinh (px/c)\quad + $$

$$\sum _{p=p_{n}}{\rm res}\;e^{pt}
\left[\frac {m(P/\alpha)c^2\cosh (pL/c)}{\sinh (pL/c)(a^2 - m^2
p^2 c^2)}\right]\sinh( px/c)\quad - $$

$$\sum _{p=p_{n}}{\rm res}\;e^{pt}
\left[\frac{(P/\alpha)pm^2c^3}{a}\;\frac {1}{(a^2 - m^2 c^2p^2)}\right]
\sinh (px/c)\quad =$$

$$u_{1} - u_{2} - u_{3} + u_{4} - u_{5},
\eqno(32)$$
where

$$u_{j} = \sum  {\rm res}\; e^{pt}\frac{A_{j}}{B_{j}}; \quad j = 1, 2, 3, 4, 5
\eqno(33)$$
and

$$A_{1} = {ac(P/\alpha) \cosh (pL/c)}\cosh (px/c);\quad
B_{1} = p\sinh (pL/c)(a^2 - m^2 p^2 c^2)\eqno(34)$$

$$A_{2} = (P/\alpha)mc^2\cosh (px/c); \quad
B_{2} = (a^2 - m^2 p^2 c^2)\eqno(35)$$

$$\quad A_{3} = (P/\alpha)c\sinh(px/c); \quad B_{3} = ap  \eqno(36)$$

$$A_{4} = (P/\alpha)mc^2\cosh (pL/c)\sinh (px/c); \quad
B_{4} = \sinh (pL/c)(a^2 - m^2 p^2 c^2)\eqno(37)$$

$$A_{5} = (P/\alpha)pm^2c^3\sinh (px/c); \quad
B_{5} = a(a^2 - m^2 p^2 c^2).\eqno(38)$$

We know from the theory of the complex functions that if the pole of
some function $f(z)/g(z)$ is simple and it is at point $a$,
then the residuum is as follows [2]:

$${\rm residuum}  = \frac{f(a)}{g'(a)}.\eqno(39)$$

If the pole at point $a$ of the function $f(z)$ is multiply of the
order $m$, then the residuum is defined as follows:

$${\rm residuum}  =  \frac{1}{(m-1)!} \lim_{z\to a}\frac{d^{m-1}}{dz^{m-1}}
\left[(z-a)^m f(z) \right].\eqno(40)$$

Let us first determined the function

$$u_{1} = \sum {\rm res}\; e^{pt}\frac{A_{1}}{B_{1}}.\eqno(41)$$

Poles of $B_{1}$ are
at points $p=0$, this is pole of the order 2, $p = + a/mc, p = -a/mc$ and
$p_{n} = + i\pi nc/L, p_{n} = -i\pi nc/L, n = 1, 2, 3,  ...$. So, the
function $u_{1}$ is as follows:

$$u_{1} =  \frac{(P/\alpha)c^2}{La}t -
 \frac{(P/\alpha)c}{a}\cosh\left(\frac{aL}{mc^2}\right)
\cosh\left(\frac{ax}{mc^2}\right)\sinh\left(\frac{at}{mc}\right)
\quad  +  $$

$$\sum_{n=1}^{n= \infty} \frac{2a(P/\alpha)c}{\pi n}
\frac{L^2}{a^2 L^2 + m^2\pi^2
n^2 c^4}\cos\left(\frac{\pi nx}{L}\right)
\sin\left(\frac{\pi n c t}{L}\right).
\eqno(42)$$.

For the function $u_{2}$ we get:
$$u_{2} =
\left(-\frac{(P/\alpha)c}{a}\right)\sinh\left(\frac{at}{mc}\right)\cosh
\left(\frac{ax}{mc^2}\right).\eqno(43)$$

$$u_{3} = 0 .\eqno(44)$$

For $u_{4}$ and $u_{5}$ we get:

$$u_{4} =
\left(-\frac{(P/\alpha)c}{a}\right)
\coth\left(\frac{aL}{mc^2}\right)\sinh\left(\frac{ax}{mc^2}\right)
\sinh\left(\frac{at}{mc}\right)\eqno(45)$$

$$u_{5} =
\left(-\frac{(P/\alpha)c}{a}\right)
\sinh\left(\frac{ax}{mc^2}\right)
\sinh\left(\frac{at}{mc}\right)\eqno(46)$$

The dimensionality of $u$ is [$u$] = m and $u(x,0) = 0$. The momentum
of a left particle $p = mu(0,t)$, or right particle  $p = mu(L,t)$ is
not conserved. Only the total momentum of a system is conserved.

\section{Discussion}

Our problem is the modification of some problems involved in the
textbooks on mathematical physics. However, our approach is
pedagogically original in
the sense that we use the initial force of a delta-function form
to show the internal motion of the string, or, rod. The delta-function
form of electromagnetic pulse was used by author in [6] and [7] to
discuss the quantum motion of an electron in the laser pulse.
We have considered here the real strings and rods in the real
space and we do not use extra-dimensions and unrealistic strings.
The M-dimensional geometrical object cannot be realized in
N-dimensional space for M $>$ N [8].
The mathematical theory of unrealistic strings is well known
as the string theory in particle physics.
Our problem with the real strings and rods
can be generalized for the two-dimensional and three-dimensional
situation. It can be also generalized to the
situation with the dissipation of waves in the strings and rods. In
this case it is necessary to write the wave equation with the
dissipative term and then to solve this problem ``ab initio''.

While we have solved the problem for the situation where the pulse was
generated by the force of the delta-form, we give some  general
ideas following from the wave equation. It is well known that
the solution of this equation is in general in the form [9]:

$$f\left(t - \frac{x}{c}\right);  \quad g\left(t + \frac{x}{c}\right)
\eqno(47)$$
where functions $f, g$ are general. It means it involves also the
function of he delta-form. For the wave propagating from the left side
to the right side,  we take function $f$. The corresponding tension in the
rod is

$$T =  ESu_{x}(x, t) = ESf'\left(t - \frac{x}{c}\right)
\left(\frac{-1}{c}\right).\eqno(48)$$

We easily see that $T(x = 0, t= 0) = T(x = L,t = L/c )$,  and it means
that when the pulse force is created at the left end of the rod then
it propagates in the rod and after time $L/c$
it is localized in the right end of the rod.

However, we have seen that the pulse force generated in the
system with the massive ends
of strings or rods develops in time according to laws of the mathematical
physics of strings and rods and cannot be intuitively predicted.
Only rigorous solution of the dynamics of the system can give the
answer on the real motion of tension in the string.

There is no information in the Newton ``Principia mathematica '' [1] and
 in any textbook on mechanics on the central collision of two particles
 where the force is mediated by string or rod.
Similarly, there is not the
 solution of our problem in the famous monograph by Pars [10].
So, This is the missing problem in the textbooks on mechanics.

The propagation of a pulse in one direction was confirmed
 experimentally by author using
 the heavy elastic rod (the segment of a rail). The delta-form force (tension) was
 generated approximately by the strike of hammer.
The experiment was performed as the table experiment and
 it can be repeated by any theorist.

The proposed model with the string with massive ends
can be also related in the modified form to the problem of
the radial motion of quarks bound by a strings, and
used to calculate the excited
states of such system. The resent  analysis of such problem
was performed by Lambiase and Nesterenko [11] and Nesterenko and
Pirozhenko [12], and others. So, can we hope  that our approach
and their approach will be unified to generate the new revolution
of the string theory of matter and space-time? Why not?

\end{document}